\documentclass[12pt]{article}
\usepackage{epsfig,amsmath,amsfonts,amssymb}
\def\eq#1\en{\begin{equation}#1\end{equation}}  
\def\eqa#1\ena{\begin{align}#1\end{align}}
\def\eqg#1\eng{\begin{gather}#1\end{gather}}
\newcommand{\lb}[1]{\label{e:#1}}
\newcommand{\rlb}[1]{\eqref{e:#1}}     
\newcommand{\bn}{\bigskip\noindent}
\newcommand{\ret}{\notag\\}
\newcommand{\qedm}{\rule{1.5mm}{3mm}}
\newcommand{\bA}{{\bar{A}}}
\newcommand{\hA}{\hat{A}}
\newcommand{\hH}{\hat{H}}
\newcommand{\hN}{\hat{N}}
\newcommand{\hP}{\hat{P}}
\newcommand{\hV}{\hH_\mathrm{int}}
\newcommand{\Amc}{\langle\hat{A}\rangle_\mathrm{mc}}
\newcommand{\DE}{\mathit{\Delta}E}
\newcommand{\bkt}[1]{\bigl\langle#1\bigr\rangle}
\newcommand{\sbkt}[1]{\langle#1\rangle}
\newcommand{\sqbkt}[1]{\sbkt{#1}_c}
\newcommand{\calG}{\mathcal{G}}
\newcommand{\calB}{\mathcal{B}}
\newcommand{\ez}{\epsilon_0}
\newcommand{\calH}{\mathcal{H}}
\newcommand{\ent}{\epsilon_n(\theta)}
\newcommand{\oL}{\{1,2,\ldots,L\}}
\newcommand{\Pv}{\Phi_\mathrm{vac}}
\newcommand{\PG}{\Phi_\Gamma}
\newcommand{\HE}{\calH_{E,\DE}}
\newcommand{\sumtwo}[2]%
{\mathop{\sum_{#1}}_{#2}}
\newcommand{\bssigma}{\boldsymbol{\sigma}}
\makeatletter
\@addtoreset{equation}{section}
\makeatother

\makeatletter
\long\def\@makecaption#1#2{{\small
\advance\leftskip1cm
\advance\rightskip1cm
\vskip\abovecaptionskip
\sbox\@tempboxa{#1: #2}%
\ifdim \wd\@tempboxa >\hsize
 #1: #2\par
\else
\global \@minipagefalse
\hb@xt@\hsize{\hfil\box\@tempboxa\hfil}%
\fi
\vskip\belowcaptionskip}}
\makeatother
\begin{document}
\noindent
{\bf{\large
The approach to thermal equilibrium and ``thermodynamic normality''}
--- 
An observation based on the works by Goldstein, Lebowitz, Mastrodonato, Tumulka, and Zangh\`\i\ in 2009, and by von Neumann in 1929
}
\setcounter{footnote}{1}
\footnotetext{Archived as {\tt 	arXiv:1003.5424}.}

\begin{flushright}
Hal Tasaki\footnote{
Department of Physics, Gakushuin University, Mejiro, Toshima-ku, Tokyo 171-8588, Japan, {\tt hal.tasaki\makeatletter@\makeatother gakushuin.ac.jp}
}
\end{flushright}

\bigskip
\begin{abstract}
We treat the problem of the approach to thermal equilibrium by only resorting to quantum dynamics of an isolated macroscopic system.
Inspired by the two important works in 2009 and in 1929, we have noted that a condition we call ``thermodynamic normality'' for a macroscopic observable guarantees the approach to equilibrium (in the sense that a measurement of the observable at time $t$ almost certainly yields a result close to the corresponding microcanonical average for a sufficiently long and typical $t$).
A crucial point is that we  make no assumptions on the initial state of the system, except that its energy is distributed close to a certain macroscopic value.

We also present three (rather artificial) models in which the thermodynamic normality can be established, thus providing concrete examples in which the approach to equilibrium is rigorously justified.

Note that this kind of results which hold for {\em any}\/ initial state are never possible in classical systems.
We are thus dealing with a mechanism which is peculiar to quantum systems.

The present note is written in a self-contained (and hopefully readable) manner.
It only requires basic knowledge in quantum physics and equilibrium statistical mechanics.
\end{abstract}

\tableofcontents

\section{Background and main results}

\subsection{Setup, background, and motivation}
\label{s:setup}

\subsubsection*{Setup}
Consider a finite but macroscopic quantum mechanical system which is completely isolated from the outside world.
The system is fully described by its Hamiltonian $\hH$.
For $\alpha=1,2,\ldots$, we denote by $E_\alpha$ and $\psi_\alpha$ the eigenvalue and the corresponding normalized eigenstate of $\hH$.
We assume that the eigenvalues are non-degenerate, i.e., $E_\alpha\ne E_{\alpha'}$ whenever $\alpha\ne\alpha'$.

We want to discuss the thermal equilibrium state of the system when it has a macroscopic energy $E$.
The standard (and empirically justified) procedure is to invoke the ``principle of equal probability'', and declare that all the microscopic states within the energy range\footnote{
The choice of this range is rather arbitrary.
The only requirements are that (i)~macroscopic properties of the system remain almost the same through the range, and (ii)~there are huge number of energy eigenstates within the range.
It is misleading to imagine that $\DE$ is determined by physical processes such as the measurement of the energy. 
} $[E,E+\DE]$ contribute equally to the equilibrium state.
Here $\DE$ is a small, but still macroscopic energy.

To be precise let $\hA$ be any (macroscopic) observable of the system.
Its microcanonical average at energy $E$ is defined by\footnote{
When $A$ is defined in terms of $B$, we write $A:=B$ or $B=:A$.
}
\eq
\Amc:=
\biggl(\sumtwo{\alpha}{(E_\alpha\in[E,E+\DE])}1\biggr)^{-1}
\sumtwo{\alpha}{(E_\alpha\in[E,E+\DE])}\bkt{\psi_\alpha,\hA\,\psi_\alpha}
=\frac{\mathrm{Tr}_{\HE}[\hA]}{\mathrm{Tr}_{\HE}[1]}
\lb{MC}
\en
where we denoted by $\HE$ the Hilbert space spanned by all the energy eigenstates $\psi_\alpha$ such that $E_\alpha\in[E,E+\DE]$, and $\mathrm{Tr}_{\HE}[\cdots]$ denotes the trace over this space.

Our goal is to justify the use of the microcanonical average \rlb{MC} by  resorting only to quantum mechanics.

\subsubsection*{Initial state and time evolution}
Suppose that at initial time $t=0$, the system was in a quantum mechanical pure state $\varphi(0)$.
We assume that the initial pure state has energy distributed near $E$.
For simplicity we let $\varphi(0)$ be an arbitrary normalized state in the above defined space $\HE$.
Of course $\HE$ contains many states which are radically far from equilibrium.

Now write this initial state as
\eq
\varphi(0)=\sum_\alpha c_\alpha\,\psi_\alpha
\lb{p0}
\en
where we have $c_\alpha=0$ if $E_\alpha\not\in[E,E+\DE]$, and $\sum_\alpha|c_\alpha|^2=1$.
The time evolution of this state is given by
\eq
\varphi(t)=\sum_\alpha c_\alpha\,e^{-i\,E_\alpha\,t}\,\psi_\alpha
\lb{pt}
\en
for any $t>0$.
Our expectation is that for sufficiently long and typical time $t$, macroscopic properties of the pure state $\varphi(t)$ can be described by the microcanonical average \rlb{MC} in the sense that a measurement of $\hA$ almost certainly yields a result close to $\Amc$.
This expectation is based on the standpoint called ``individualist'' point of view in \cite{GLTZ10}. 
See \cite{GLTZ10} for a list of related works\footnote{
It is sometimes argued that one should consider a system coupled to the outside world because no systems are perfectly isolated in reality.
Let us emphasize, however, that it is perfectly reasonable to consider an isolated system as an idealization.
After convincing ourselves that an isolated macroscopic system has a tendency to reach its equilibrium, we can start thinking about why and how the coupling to the outside world does not destroy this tendency.
}.

\subsubsection*{Basic picture}
Let us briefly discuss the idea behind the above expectation that $\varphi(t)$ should be described by the microcanonical average.
See \cite{L07} (and \cite{TasakiSM}, if the reader can read Japanese) for more details.

\begin{figure}
\centerline{\epsfig{file=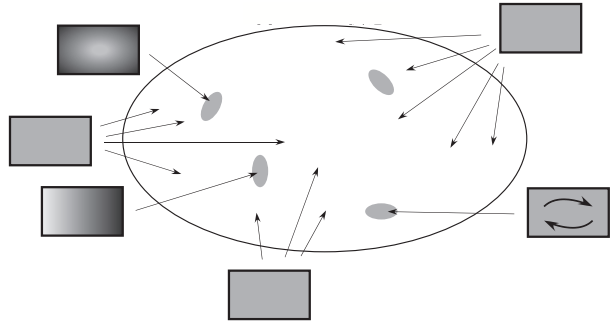,width=8cm}}
\caption[dummy]{
The big elliptic region represents the Hilbert space $\HE$, the space of states that have a macroscopic energy $E$.
Overwhelming majority of the states look almost the same from macroscopic points of view, and they exhibit properties of the corresponding equilibrium states.
Exceptional states which belong to very small shaded regions have macroscopic properties different from equilibrium.
Note that this is a very crude picture of the Hilbert space with extremely high dimension.
Also note that the exceptional regions must be much smaller than depicted.
The figure is taken from \protect{\cite{TasakiSM}}.
}
\label{f:AS}
\end{figure}

We first argue that the universal applicability of equilibrium statistical mechanics strongly suggests that {\em (i)~overwhelming majority of sates in the space $\HE$ look almost similar from macroscopic points of view}\/, and {\em (ii) ``equilibrium properties'' are nothing but the common properties that are shared by these majority of states}\/.
In other words, most of the states in $\HE$ can be regarded (from macroscopic points of view) as representatives of the equilibrium state.
The remaining exceptional minorities correspond to varieties of nonequilibrium states.
See Figure~\ref{f:AS}.
It is then evident that by averaging over all the states in $\HE$ as in \rlb{MC}, one can extract the properties of the majority since exceptional contributions are so minor and averaged out.

\begin{figure}
\centerline{\epsfig{file=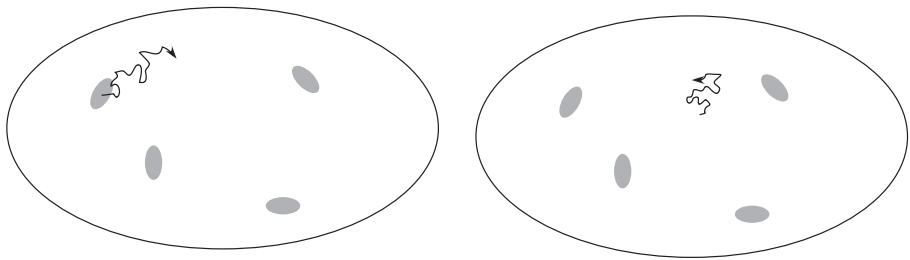,width=12cm}}
\caption[dummy]{
The basic mechanism of the approach to equilibrium based on the picture as in Figure~\protect{\ref{f:AS}}.
Left:~If the state was initially in one of the exceptional regions, it eventually moves out of the  region and evolves into the overwhelming majority of typical states.
Right:~If the state was initially typical, it rarely wanders into the exceptional regions, and keeps being typical.
The figure is taken from \protect{\cite{TasakiSM}}.
}
\label{f:IR}
\end{figure}

Once accepting this picture, we can naturally see why $\varphi(t)$, in the long run, should approach equilibrium.
Suppose that the initial state $\varphi(0)$ describes a physical situation which is very far from equilibrium, e.g., two bodies at different temperatures in contact with each other\footnote{
We treat (an artificial version of) this problem in section~\ref{s:Two}.
}.
The initial state $\varphi(0)$ certainly is an ``exceptional'' state which does not belong to the overwhelming majority.
Obviously the coefficients $c_\alpha$ in \rlb{p0} must be chosen with an extreme care and precision in order to realized such an exceptional state.
As $t$ grows, each coefficient gets individual phase factor as in \rlb{pt}, and the very delicate balance of the coefficients realized at $t=0$ will soon be lost.
It is quite likely that, after a sufficiently long time, the state $\varphi(t)$ is no longer exceptional, and belongs to the overwhelming majority of equilibrium states.
See Figure~\ref{f:IR}.

\subsubsection*{Older works\footnote{
Note that this part of the note is a rather personal account which places the present observation in a broader context.
Please refer to \cite{GLTZ10} and references therein for a more balanced view on what have been done.
}}
If one wants to justify the above expectation that $\varphi(t)$ becomes an equilibrium state, a natural (and well-known) starting point is to consider a long-time average.

Let $\hA$ be a macroscopic observable.
Its quantum mechanical expectation value in the state $\varphi(t)$ is
\eq
\bkt{\varphi(t),\hA\,\varphi(t)}=
\sum_{\alpha,\alpha'}
\overline{c_\alpha}\,c_{\alpha'}\,e^{i\,(E_\alpha-E_{\alpha'})\,t}\,
\bkt{\psi_\alpha,\hA\,\psi_{\alpha'}},
\lb{At}
\en
where we used \rlb{pt}.
Since we assumed that the energy eigenvalues are nondegenerate, all the terms with $\alpha\ne\alpha'$ exhibit oscillations.
By averaging over infinite amount of time, the oscillations simply average out, and one finds
\eq
\lim_{T\uparrow\infty}\frac{1}{T}\int_0^Tdt\,\bkt{\varphi(t),\hA\,\varphi(t)}
=\sum_\alpha|c_\alpha|^2\,\bkt{\psi_\alpha,\hA\,\psi_{\alpha}},
\lb{ATinf}
\en
which is remarkably similar to the microcanonical average \rlb{MC}.

Then the challenge was to show that the quantum mechanical expectation value  (without the time average) is close to the microcanonical average.
More precisely, to show that
\eq
\bkt{\varphi(t),\hA\,\varphi(t)}\simeq \Amc\quad
\text{for most $t$}
\lb{oldgoal}
\en
provided that we consider long enough time interval.
This, if done, may be regarded as a justification of the microcanonical average\footnote{
To be more precise, one further has to assume that quantum mechanical fluctuation of $\hA$ in $\varphi(t)$ is small.
}.

In \cite{Tasaki98}, the temporal fluctuation of $\bkt{\varphi(t),\hA\,\varphi(t)}$ was evaluated using the standard Chebyshev inequality argument\footnote{
It was also assumed that $E_\alpha-E_{\alpha'}=E_{\alpha''}-E_{\alpha'''}\ne0$ implies $\alpha=\alpha''$, $\alpha'=\alpha'''$.
This is called the non-resonance condition.
}.
It was shown that the desired \rlb{oldgoal} is justified under the assumption that each coefficient $c_\alpha$ is small\footnote{
\cite{Tasaki98} discusses a setup leading to the canonical ensemble, but the microcanonical setting is treated in exactly the same (indeed easier) manner.
See also \cite{unpub2} where we treated a microcanonical problem.
}.
In other words, the initial state $\varphi(0)$ should have energy distributed around $E$, but not too sharply.
A different criterion of the same nature was recently obtained in \cite{Reimann}.
See section~\ref{s:most} for a new result of the same philosophy.

It is a very difficult question whether we should be satisfied with this type of ``derivation of statistical mechanics from quantum mechanics.''
The most delicate point is the restriction on the allowed initial state $\varphi(0)$.
It seems to be true that excluded initial states (which have sharp peaks at some energies) are ``rare'', and the practical ``chance'' of choosing such initial states in actual world can be safely neglected.
But in what sense are they ``rare'' or do they have negligible ``chances''?
We are dealing with a preparation of a pure state, where we do not see any  natural concept of probabilities.

\subsubsection*{A new (but in fact rather old) thought}
In \cite{GLMTZ09b}, Goldstein, Lebowitz, Mastrodonato, Tumulka, and Zangh\`\i\   proposed an abstract formulation of equilibrium states in a macroscopic system.
They were able to prove that, in a typical system, {\em any}\/ initial state $\varphi(0)\in\HE$ will eventually evolve into equilibrium.
In other words the approach to equilibrium is guaranteed  by only assuming that the initial state has energy distributed around $E$.
Note that this assumption on the range of energy is almost mandatory since we never observe a linear combination of multiple states with macroscopically different energies\footnote{
It is possible that such a state (Schr\"odinger's cat) is inhibited (or destroyed) by interaction of the system with the outside world.
But this is a totally different topic (for the moment).
}.

Such a result with basically {\em no}\/ restrictions on the initial state has a great advantage, at least from a conceptual point of view.
With such a statement, we no longer have to worry about the ``chance'' of choosing an exceptional initial state.
It should be stressed at this moment that, in a classical system, it is impossible to have a similar statement which holds for {\em any}\/ initial states.
In any classical dynamical system which shows thermalization, there exists a set of ``exceptional'' initial states which do not relax to equilibrium.
The set of exceptional states usually have a vanishing (Lebesgue) measure, and one never chooses such a state provided that the initial state is sampled according to the microcanonical ensemble\footnote{
Although this is usually regarded as a satisfactory justification of the equilibrium statistical mechanics, it should be noted that here one is invoking a kind of circular reasoning.
There seems to be no a priori reason (based solely on classical mechanics) to use the microcanonical (or the Lebesgue) measure when one samples the initial state.
It seems possible that the uncertainty principle from quantum mechanics provides a reason.
}.

Interestingly (and surprisingly), back in 1929, a result of the same nature as that of Goldstein, Lebowitz, Mastrodonato, Tumulka, and Zangh\`\i\   was already derived by von Neumann \cite{vN29}.
He treated the problem of justifying statistical mechanics by using quantum mechanics, and stated a theorem which holds for {\em any}\/ initial states.
We highly recommend the readers to go through \cite{GLTZ10}, which is a very well written commentary by Goldstein, Lebowitz, Tumulka, and Zangh\`\i\ on this important (and long forgotten) work by von Neumann.

We will discuss the result of Goldstein, Lebowitz, Mastrodonato, Tumulka, and Zangh\`\i\  \cite{GLMTZ09b} agin in section \ref{s:rel}.

\subsection{Definition of ``thermodynamic normality'' and the main theorem}
\label{s:main}
We are now ready to describe our simple observation inspired by  the results of Goldstein, Lebowitz, Mastrodonato, Tumulka, and Zangh\`\i\  \cite{GLMTZ09b} and also of von Neumann \cite{vN29}.
Like these authors, we insist on deriving a result that is valid for {\em any}\/ choice of initial state $\varphi(0)$ whose energy is distributed around $E$.
We use the same notation as in section~\ref{s:setup}.

The key of the observation is the following definition of {\em thermodynamic normality}\/ of a macroscopic observable\footnote{
\label{fn:normal}
This terminology was suggested to the author by Joel Lebowitz and Sheldon Goldstein.
The adjective ``thermodynamic'' stresses that we are dealing with a macroscopic observable in a macroscopic system where thermodynamic description is meaningful. 
We remark, however, that the thermodynamic normality is neither a sufficient nor a necessary condition for the system to exhibit thermodynamic behaviors.
Also note that, in \cite{GLMTZ09b}, the term ``normal'' is used to indicate a more detailed property.
}.

\bn
{\bf Definition:} We say that a macroscopic observable $\hA$ is {\em thermodynamically normal}\/ with respect to $\hH$ in the interval $[E,E+\DE]$ if 
\eq
\bkt{\psi_\alpha,(\hA-\Amc)^2\,\psi_\alpha}\le \bA^2\,\zeta
\lb{normal}
\en
holds for any $\alpha$ such that $E_\alpha\in[E,E+\DE]$, where $\zeta>0$ is a small constant.
Here constant $\bA$ (which may depend on $E$) is a (rather arbitrarily chosen) typical magnitude of $\hA$.
\bigskip

Note that by averaging the above \rlb{normal} over all $\alpha$ such that $E_\alpha\in[E,E+\DE]$, one gets
\eq
\bkt{(\hA-\Amc)^2}_\mathrm{mc}\le \bA^2\,\zeta
\lb{normalMC}
\en
which simply says that the fluctuation of $\hA$ is small in the microcanonical ensemble.
This property is believed to hold quite generally, and may be proved rigorously under suitable concrete settings (although the proof may not be easy).

By the thermodynamic normality  \rlb{normal}, on the other hand, we are requiring that, as far as the observable $\hA$ is concerned, each eigenstate $\psi_\alpha$ in the energy range $[E,E+\DE]$ behaves almost as the equilibrium state.
This is indeed a very strong requirement, which may or may not be true for general macroscopic systems.
In section~\ref{s:examples}, we discuss (rather artificial) examples where the thermodynamic normality can be justified.
See section~\ref{s:discussion} for further discussions.

The following main theorem is indeed a trivial consequence of the  above definition.

\bn
{\bf Theorem:}
Let $\eta$ and $\delta$ be small quantities, and assume that  $\hA$ satisfies the thermodynamic normality \rlb{normal} with $\zeta=2\eta^2\,\delta$.
Then for {\em any}\/ initial state $\varphi(0)\in\HE$, one can find a (sufficiently long) time $T$, and a ``good'' subset $\calG\subset[0,T]$ with\footnote{
$|\calG|$ denotes the Lebesgue measure (the total length) of $\calG$.
} $|\calG|/T\ge1-\delta$, such that
\eq
\bkt{\varphi(t),(\hA-\Amc)^2\,\varphi(t)}\le (\bA\,\eta)^2
\lb{good}
\en
holds for any $t\in\calG$.

\bigskip

Roughly speaking, \rlb{good} says that for any $t\in\calG$, we have
\eq
\hA=\Amc+O(\bA\,\eta)\quad\text{in $\varphi(t)$.}
\en
Thus if one measures $\hA$ at $t$, the outcome must be close to $\Amc$ with probability very close to 1.
Moreover the set $\calG$ of such ``good'' $t$ occupies most of the time interval $[0,T]$.
Therefore, the result of quantum mechanical measurement of $\hA$ after a long enough time is almost certainly almost identical to the microcanonical prediction $\Amc$.
This justifies the use of the microcanonical average\footnote{
The present formulation automatically covers the canonical average if one regards the system as consisting of a subsystem and a heat bath.
When the subsystem itself is macroscopic, a macroscopic quantity $\hA$ of the subsystem exhibits negligible fluctuation.
}.

Note that the ``good'' set $\calG$ inevitably contains some intermittent vacancies.
A perfect settlement to equilibrium is not possible because quantum dynamics is  quasi-periodic.
The present consideration does not provide any information on how large $T$ should be to see the approach to equilibrium\footnote{
It may be possible that a completely isolated quantum system sometimes requires quite a long time to relax to the equilibrium.
}.

Let us be more precise about the above claim about the measurement.
Let $\hP_a$ be the projection operator onto the eigenspace corresponding to the eigenvalue $a$ of $\hA$.
Then $p_a(t):=\bkt{\varphi(t),\hP_a,\varphi(t)}$ is the probability that the outcome of the measurement of $\hA$ is equal to $a$.
The inequality \rlb{good} implies
\eq
\sum_a(a-\Amc)^2\,p_a(t)\le (\bA\,\eta)^2,
\en
and hence for any $s>0$, one has\footnote{
The argument used here is that for the Chebyshev's inequality.
}
\eq
\mathrm{Prob}_t\Bigl[|a-\Amc|\ge \bA\,s\Bigr]
:=\sumtwo{a}{(|a-\Amc|\ge \bA\,s)}p_a(t)
\le\left(\frac{\eta}{s}\right)^2
\en
By taking   $s$ so that  $\eta\ll s\ll 1$, the right-hand side becomes small, which means that the (quantum mechanical) probability of observing a value far from $\Amc$ is small.

\bigskip
\bn
{\em Proof of Theorem:}\/
From \rlb{pt}, we have
\eq
\bkt{\varphi(t),(\hA-\Amc)^2\,\varphi(t)}=\sum_{\alpha,\alpha'}
\overline{c_\alpha}\,c_{\alpha'}\,e^{i\,(E_\alpha-E_{\alpha'})\,t}\,
\bkt{\psi_\alpha,(\hA-\Amc)^2\,\psi_{\alpha'}}.
\lb{ext}
\en
Since the energy eigenvalues are nondegenerate, the long time average of \rlb{ext} becomes
\eq
\lim_{T\uparrow\infty}\frac{1}{T}\int_0^Tdt\,\bkt{\varphi(t),(\hA-\Amc)^2\,\varphi(t)}
=\sum_\alpha|c_\alpha|^2\,\bkt{\psi_\alpha,(\hA-\Amc)^2\,\psi_{\alpha}}
\le2(\bA\,\eta)^2\,\delta, 
\lb{Tinf}
\en
where we noted that $\sum_\alpha|c_\alpha|^2=1$, and used the thermodynamic normality \rlb{normal} with $\zeta=2\eta^2\,\delta$.
Since we have the bound \rlb{Tinf} for $T\uparrow\infty$, there exists $T>0$ such that
\eq
\frac{1}{T}\int_0^Tdt\,\bkt{\varphi(t),(\hA-\Amc)^2\,\varphi(t)}
\le(\bA\,\eta)^2\,\delta, 
\lb{Tfin}
\en
Define the ``bad set'', on which $\hA$ deviates considerably from $\Amc$, by
\eq
\calB:=\Bigl\{t\in[0,T]\,\Bigl|\,\bkt{\varphi(t),(\hA-\Amc)^2\,\varphi(t)} > (\bA\,\eta)^2\Bigr\}.
\en
Note that $\calG=[0,T]\backslash\calB$, and $|\calB|=T-|\calG|$.
Then we have
\eq
\frac{1}{T}\int_0^Tdt\,\bkt{\varphi(t),(\hA-\Amc)^2\,\varphi(t)}
\ge\frac{1}{T}\int_{t\in\calB}dt\,(\bA\,\eta)^2=(\bA\,\eta)^2\frac{|\calB|}{T}=(\bA\,\eta)^2\,\Bigl(1-\frac{|\calG|}{T}\Bigr),
\en
which, with \rlb{Tfin}, implies the desired bound on $|\calG|/T$.~\qedm

\subsection{A result which holds for most initial states}
\label{s:most}
If one is satisfied with a result which holds for {\em most} initial state $\varphi(0)$ from $\HE$, it is enough to make much weaker assumptions than the thermodynamic normality.
Although this is beside the main point of the present note, let us briefly describe some facts for completeness.

By examining the above proof, one easily finds that the inequality in \rlb{Tinf} is the key for the subsequent estimates.
Thus we readily find

\bn
{\bf Proposition 1:}
We do not assume the thermodynamic normality.
If the initial state $\varphi(0)=\sum_\alpha c_\alpha\psi_\alpha\in\HE$ is such that
\eq
\sum_\alpha|c_\alpha|^2\,\bkt{\psi_\alpha,(\hA-\Amc)^2\,\psi_{\alpha}}
\le2(\bA\,\eta)^2\,\delta
\lb{cd}
\en
then the conclusion of the main theorem holds.

\bigskip

We can also show that the condition \rlb{cd} is indeed satisfied for a typical state in $\HE$ by assuming that the observable $\hA$ exhibits small fluctuation in the microcanonical ensemble as in \rlb{normalMC}.

\bn
{\bf Proposition 2:}
We do not assume the thermodynamic normality.
Suppose that the inequality \rlb{normalMC} is valid with $\zeta=2\,\eta^2\,\delta\,\nu$, where $\nu$ is a small constant.
If one samples a normalized state $\varphi(0)$ uniformly\footnote{
We do not claim that the uniform sampling is physically realistic.
The present proposition simply tells us that there are many $\varphi(0)$ which satisfies the condition \rlb{cd}.
} from the space $\HE$, then $\varphi(0)$ satisfies the condition \rlb{cd} with the probability not less than $1-\nu$.

\bn
{\em Proof:}
Write $I:=\{\alpha\,|\,E_\alpha\in[E,E+\DE]\}$, and denote by $\sqbkt{\cdots}$ the uniform average over $(c_\alpha)_{\alpha\in I}$ with $c_\alpha\in\mathbb{C}$ and $\sum_{\alpha\in I}|c_\alpha|^2=1$.
Note that the symmetry implies $\sqbkt{\overline{c_\alpha}\,c_{\alpha'}}=\delta_{\alpha,\alpha'}/|I|$, where $|I|$ denotes the number of elements in $I$.

Write for simplicity $f_\alpha:=\bkt{\psi_\alpha,(\hA-\Amc)^2\,\psi_{\alpha}}$.
Then the assumption of the proposition reads
\eq
\sqbkt{{\textstyle\sum_\alpha|c_\alpha|^2f_\alpha}}=\frac{1}{|I|}\sum_\alpha f_\alpha=\bkt{(\hA-\Amc)^2}_\mathrm{mc}\le 2\,(\bA\,\eta)^2\,\delta\,\nu
\lb{cf}
\en
By denoting $\chi[\text{true}]=1$, $\chi[\text{false}]=0$, we see
\eq
\textstyle\sqbkt{\sum_\alpha|c_\alpha|^2f_\alpha}\ge 2\,(\bA\,\eta)^2\,\delta\,\sqbkt{\,\chi[\,\sum_\alpha|c_\alpha|^2f_\alpha\ge2\,(\bA\,\eta)^2\,\delta\,]\,}
\en
which with \rlb{cf} implies
\eq
\textstyle
\mathrm{Prob}[\,\sum_\alpha|c_\alpha|^2f_\alpha\ge2\,(\bA\,\eta)^2\,\delta\,]
:=\sqbkt{\,\chi[\,\sum_\alpha|c_\alpha|^2f_\alpha\ge2\,(\bA\,\eta)^2\,\delta\,]\,}
\le\nu,
\en
which is the desired bound.~\qedm

\subsection{Relation to the result of Goldstein, Lebowitz, Mastrodonato, Tumulka, and Zangh\`\i}
\label{s:rel}

Let us briefly discuss the results of Goldstein, Lebowitz, Mastrodonato, Tumulka, and Zangh\`\i\  \cite{GLMTZ09b}, and how our main observation in section~\ref{s:main} is related to it (indeed in a straightforward manner).

Goldstein, Lebowitz, Mastrodonato, Tumulka, and Zangh\`\i\  start by assuming that there is a subspace $\calH_\mathrm{eq}\subset\HE$ in which various macroscopic quantities take their equilibrium values.
They assume that the dimensions of the spaces satisfy
\eq
\frac{\mathrm{dim}\,\calH_\mathrm{eq}}{\mathrm{dim}\,\HE}\simeq1.
\lb{dH1}
\en
This assumption is consistent with the picture that overwhelming majority of the states in $\HE$ can be regarded as representing the equilibrium (recall  Figure~\ref{f:AS}).
Let $\hat{P}_\mathrm{eq}$ be the orthogonal projection onto the equilibrium Hilbert space $\calH_\mathrm{eq}$.

For a given equilibrium Hilbert space $\calH_\mathrm{eq}$, consider the condition
\eq
\bkt{\psi_\alpha,\hat{P}_\mathrm{eq}\,\psi_\alpha}\simeq 1
\quad\text{for all $\alpha$ such that $E_\alpha\in[E,E+\DE]$},
\lb{mynormal}
\en
which, in our terminology\footnote{
As we remarked before in the footnote \ref{fn:normal}, the authors of \cite{GLMTZ09b} use the term ``normal'' in a different manner.
}, should be called ``thermodynamic normality.''

Then the results of Goldstein, Lebowitz, Mastrodonato, Tumulka, and Zangh\`\i\ \cite{GLMTZ09b} can be summarized (in our own interpretation) as follows.

\bn
(A)~If the equilibrium subspace $\calH_\mathrm{eq}$ satisfies the thermodynamic normality \rlb{mynormal}, then {\em any}\/ initial state $\varphi(0)\in\HE$ will eventually evolve into equilibrium in the sense that
\eq
\bkt{\varphi(t),\hat{P}_\mathrm{eq}\,\varphi(t)}\simeq 1
\en
holds for most $t$ in a sufficiently long time interval.

\bn
(B)~Fix the Hamiltonian $\hH$, and choose the equilibrium space $\calH_\mathrm{eq}$ in a random manner\footnote{
In the original work \cite{GLMTZ09b}, $\calH_\mathrm{eq}$ is fixed, and $\hH$ is chosen in a random manner.
But mathematically speaking, these two formulations are the same.
}.
Then with a probability very close to 1, the equilibrium subspace $\calH_\mathrm{eq}$ satisfies  \rlb{mynormal}.
Thus the thermodynamic normality is a typical property.

\bigskip
Note that, as the authors of \cite{GLMTZ09b} themselves warn, the random choice of $\calH_\mathrm{eq}$ in (B) should not be taken too literally.
This formulation does not mean that either the space $\calH_\mathrm{eq}$ or the Hamiltonian $\hH$ is chosen according to certain stochastic rules in reality.
The typicality (B) should be understood as a strong indication that the thermodynamic normality  \rlb{mynormal} is a common property which is shared by many pairs of  $\calH_\mathrm{eq}$ and $\hH$.

The nontrivial part of this work is (B), and the derivation of (A) is rather straightforward\footnote{
Of course, to find the correct claim to prove is never straightforward.
}.
Nevertheless, we were very much impressed by the claim (A).
It can be read as a concrete criterion that $\calH_\mathrm{eq}$ and $\hH$ should satisfy to guarantee the approach to equilibrium for {\em any}\/ initial state.
Then a natural challenge is to take a ``constructive'' approach, i.e., to choose  concrete (and hopefully nontrivial) models and show that the criterion is actually satisfied (or show that it is not satisfied).

To treat the expectation value of the projection $\hat{P}_\mathrm{eq}$, however, is not an easy task in general.
Since the equilibrium space $\calH_\mathrm{eq}$ is determined by possible macroscopic observables, it might be easier to work directly with observables than the projection.
If one pursues this line of thought, and reinterprets the above (A) in terms of a macroscopic operator, then one gets our Theorem 1 in an almost straightforward manner.

In the next section, we present the first step of our ``constructive'' approach by proving the thermodynamic normality of some observables in three concrete models.
In the example of thermal contact treated in section~\ref{s:Two}, we can treat the projection $\hat{P}_\mathrm{eq}$ explicitly and show the property \rlb{mynormal}.
In the free fermion model treated in section~\ref{s:fermi}, on the other hand, we still do not see how to treat  $\hat{P}_\mathrm{eq}$  while we are able to show the thermodynamic normality for certain observables.

\section{Examples}
\label{s:examples}
We shall discuss three classes of examples where we can establish the thermodynamic normality of certain observables explicitly.
Thus they provide rigorous examples where the approach to thermal equilibrium from any $\varphi(0)\in\HE$ is proved rigorously.

\subsection{Independent spins under random magnetic field}
\label{s:spin}
Let us start with a trivial example of independent spins under quenched random magnetic field.
In this model  independent precession of each spin causes the ``approach to equilibrium'' for certain observables.
Although everything is trivial, it may be a good idea to look at the simple (but genuinely quantum mechanical) mechanism that realizes the relaxation-like behavior.
Interestingly, the same model also offers a counterexample to the thermodynamic normality.

\bigskip
Consider a system of $N$ independent spins with $S=1/2$.
We assume $N\gg1$.
For $\nu=x,y,z$ and $j=1,2,\ldots,N$, we denote by $\hat{S}^{(\nu)}_j$ the spin operator in the $\nu$-direction of the $j$-th spin.
We consider the Hamiltonian
\eq
\hH=-\sum_{j=1}^Nh_j\,\hat{S}^{(z)}_j
\lb{Hspin}
\en
where the quenched random magnetic field $h_j$ ($j=1,\ldots,N$) is a random quantity which are independent with each other and identically distributed according to a certain continuous probability distribution.
Since the details of the distribution is irrelevant here, one may  assume that each $h_j$ is drawn uniformly from the interval $[-h,h]$.

For each $j=1,\ldots,N$ and $\sigma=\pm1$, we denote by $\psi_j^{\sigma}$ the standard basis states of the $j$-th spin which satisfy
\eq
\hat{S}^{(z)}_j\,\psi_j^{\sigma}=\frac{\sigma}{2}\,\psi_j^{\sigma}.
\en
Let $\bssigma:=(\sigma_j)_{j=1,\ldots,N}$ be the multi-index with $\sigma_j=\pm1$, and define the corresponding $N$ spin states by
\eq
\Psi_{\bssigma}:=\bigotimes_{j=1,\ldots,N}\psi_j^{\sigma_j}.
\en
It is obvious that the state $\Psi_{\bssigma}$ is an eigenstate of $\hH$ with the eigenvalue
\eq
E_{\bssigma}:=-\sum_{j=1}^Nh_j\,\sigma_j.
\lb{Es}
\en
Since each $h_j$ is random, the energy eigenvalues \rlb{Es} are nondegenerate with probability one.

Let us take the total spin in the $x$-direction $\hat{S}_\mathrm{tot}^{(x)}=\sum_{j=1}^N\hat{S}_j^{(x)}$ as a macroscopic observable.
Then one readily finds that
\eq
\bkt{\Psi_{\bssigma},\hat{S}_\mathrm{tot}^{(x)}\,\Psi_{\bssigma}}=0,\quad
\bkt{\Psi_{\bssigma},(\hat{S}_\mathrm{tot}^{(x)})^2\,\Psi_{\bssigma}}=\frac{N}{4}
\en
for any configuration $\bssigma$.
Thus by choosing the typical magnitude of $\hat{S}_\mathrm{tot}^{(x)}$ as $N/2$, we find that the thermodynamic normality \rlb{normal} is satisfied with $\sbkt{\hat{S}_\mathrm{tot}^{(x)}}_\mathrm{mc}=0$ and $\zeta=N^{-1}\ll1$.

Note that the range of the energy is arbitrary here.
One can even start from the state
\eq
\Phi(0)=\bigotimes_{j=1,\ldots,N}\frac{\psi_j^{+1}+\psi_j^{-1}}{\sqrt{2}},
\en
which satisfies $\hat{S}_\mathrm{tot}^{(x)}\,\Phi(0)=(N/2)\,\Phi(0)$, and conclude that $\bigl|\,\sbkt{\Phi(t), \hat{S}_\mathrm{tot}^{(x)}\,\Phi(t)}\,\bigr|\ll N$ for sufficiently large and typical $t$.
As we have noted in the beginning, this ``relaxation'' is nothing but a trivial consequence of independent precession of each spins.

\bigskip
It is interesting to see what happens if we take the total spin in the $z$-direction $\hat{S}_\mathrm{tot}^{(z)}=\sum_{j=1}^N\hat{S}_j^{(z)}$ as our macroscopic observable.
Since the energy eigenvalue \rlb{Es} is the sum of (the sign factor times) the continuously distributed random quantities, it happens in general that two energy eigenvalues $E_{\bssigma}$ and $E_{\bssigma'}$ which are extremely close to each other have radically different configurations $\bssigma$ and $\bssigma'$.
As a consequence  \cite{PH}, $\sum_j\sigma_j$, which is the eigenvalue $\hat{S}_\mathrm{tot}^{(z)}$, shows a rather erratic behavior when viewed as a function of the energy eigenvalue $E_{\bssigma}$.
This implies that the macroscopic observable $\hat{S}_\mathrm{tot}^{(z)}$ is not thermodynamically normal in any energy intervals.
Indeed, the expectation value of $\hat{S}_\mathrm{tot}^{(z)}$ is independent of time, and can never approach to equilibrium.

This rather trivial example illustrates how the thermodynamic normality (as well as the tendency of approach to equilibrium) can be lost in a system with a quenched disorder.
In \cite{PH}, an interacting spin model with the Hamiltonian
\eq
\hH=J\sum_{j=1}^N\hat{\bf{S}}_j\cdot\hat{\bf{S}}_{j+1}+\sum_{j=1}^Nh_j\,\sigma_j,
\en
where $J$ is a constant and $h_j$ is uniformly distributed in $[-h,h]$, was studied numerically.
A systematic analysis suggests that the ``localization'' observed above for $J=0$ persists in a model with sufficiently small $|J|$.
In this case, the thermodynamic normality is not satisfied for a general observables, and the model lacks the ability to relax to equilibrium by itself.
For large enough $|J|$, the system enters the ``delocalized'' phase where the thermodynamic normality may hold.
See \cite{PH} and references therein for further discussions about the localization in many body quantum systems and its relation to the problem of the approach to equilibrium.

\subsection{Free fermions on a discrete chain}
\label{s:fermi}

The second example is the system of $N$ free fermions on a discrete chain of $L$ sites with the standard hopping Hamiltonian \rlb{Hfermi} (with an extra phase factor $\theta$).
We show explicitly that the particle number $\hN_\ell$ and the energy $\hH_\ell$ on a part of the lattice are normal when $N/L$ and $\ell/L$ are $O(1)$ and the system size $L$ is large.
With a further consideration on the degeneracy, we prove that both $\hN_\ell$ and $\hH_\ell$ approach their equilibrium values in the sense of the main theorem\footnote{
We thus allow {\em any}\/ initial state from $\HE$.
Although it is clear that $\HE$ contains various nonequilibrium states, it is not easy to precisely state which states are in $\HE$.
}, provided that $L$ is odd and the phase factor $\theta$ does not take one of a finite number of exceptional values.

Note that this conclusion is consistent with the well-known fact that an idealized gas never exhibits a true thermalization towards equilibrium.
We are here only looking at rather crude observables like the partial particle number and the partial energy.
These quantities can (and do) approach its equilibrium values with only ideal gas dynamics.
If we look at more sophisticated observables (such as properly coarse grained velocity distribution), then they must fail to be thermodynamically normal.

It is well known that a classical ideal gas (i.e., a classical system of a macroscopic number of non-interacting particles) also exhibits a diffusion-like behavior.
Suppose that all the particles are initially located in one end of a container.
As each particle moves in the container with its own velocity, the positional distribution of the particles gradually spreads out, and in the long run one finds that particles are distributed almost uniformly in the container.
The relaxation phenomena we see in this section is quite similar to this classical phenomenon.
It must be noted, however, that in the classical model, one needs to exclude exceptional initial states (like the one in which all the particles have the same velocity) to have the relaxation behavior.
There seems to be an essential difference with the following quantum version where we are allowed to start from {\em any}\/ initial state.

\subsubsection{Model and energy eigenstates}
We consider a system of $N$ spinless fermions on a finite periodic chain with $L$ sites.
We assume both $L$ and $N$ are large, and the density $N/L$ is of $O(1)$.

Let us denote a site on the chain as $x\in\oL$.
The Hamiltonian is
\eq
\hH=\frac{1}{2}\sum_{x=1}^L\bigl\{e^{i\theta}\,c^*_x\,c_{x+1}
+e^{-i\theta}\,c^*_{x+1}\,c_{x}\bigr\},
\lb{Hfermi}
\en
where $c_x$ and $c^*_x$ are the annihilation and the creation operators, respectively, of the fermion at site $x\in\oL$.
They satisfy the standard canonical anticommutation relations
\eq
\{c^*_x,c_y\}=\delta_{x,y},\quad\{c^*_x,c^*_y\}=0,\quad\{c_x,c_y\}=0
\lb{cac1}
\en
for any $x,y\in\oL$, where anticommutator is $\{A,B\}:=AB-BA$.
We use the periodic boundary condition $c_{L+1}=c_1$.
\rlb{Hfermi} is the standard tight-binding hopping Hamiltonian except for the phase factor $\theta\in(0,\pi/L)$, which is (rather artificially) introduced to reduce the symmetry of the system and avoid degeneracy.

For $n\in\oL$, define the creation operator for the single-particle energy eigenstate as
\eq
a^*_n:=\frac{1}{\sqrt{L}}\sum_{x=1}^L\exp\Bigl[i\,\frac{2\pi\,n}{L}\,x\Bigr]\,c^*_x,
\lb{as}
\en
which again satisfy the canonical anticommutation relations
\eq
\{a^*_n,a_{n'}\}=\delta_{n,n'},\quad\{a^*_n,a^*_{n'}\}=0,\quad\{a_n,a_{n'}\}=0,
\lb{cac2}
\en
for any $n,n'\in\oL$.
The fact that $a^*_n$ creates a single-particle energy eigenstate is encoded in the commutation relation
\eq
[\hH,a^*_n]=\epsilon_n\,a^*_n
\en
with the corresponding single-particle energy eigenvalue
\eq
\epsilon_n:=\cos\Bigl(\frac{2\pi\,n}{L}+\theta\Bigr).
\en

Fix the particle number $N$.
Let $\Pv$ be the vacuum state with no  particles on the chain, which satisfies $c_x\,\Pv=0$ for any $x\in\oL$.
For each subset $\Gamma\subset\oL$ with $|\Gamma|=N$, one has an $N$ particle energy eigenstate\footnote{
The product of $a^*_n$ is ordered according to the order of $n$.
}
\eq
\PG=\Bigl(\prod_{n\in\Gamma}a^*_n\Bigr)\Pv,
\lb{NG}
\en
with the energy eigenvalue
\eq
E_\Gamma=\sum_{n\in\Gamma}\epsilon_n.
\lb{EG}
\en

The energy eigenvalues \rlb{EG} may be degenerate.
It can be shown that by choosing $L$ to be a prime number, the energy eigenvalues are nondegenerate for most choice of $\theta\in(0,\pi/L)$ except for a finite number of exceptional points.
In this case we can apply our main theorem as it is (provided that the thermodynamic normality is guaranteed).

It is of course somewhat absurd to limit ourselves to chains whose length is exactly equal to a prime number.
In fact, in order to prove the desired conclusion about the approach to equilibrium, it is enough to assume that $L$ is odd and choose suitable $\theta$ except for a finite number of exceptional points.
See section~\ref{s:oddL}.

Although we agree that these fine-tuning concerning the degeneracy may not sound quite physical, we do not regard this as a serious problem.
Technically speaking, it is obvious that the degeneracy can always be lifted by using various ``dirty'' tricks, e.g., by adding extremely small random on-site potential energy.
It is our optimistic hope that the degeneracy (in a clean system) is an accidental phenomenon in a non-interacting systems, and may be naturally avoided in fully interacting systems unless there are special symmetries.

\subsubsection{Thermodynamic normality}
\label{s:fermiN}
Let us discuss the thermodynamic normality \rlb{normal} for the following two observables.

Fix $\ell<L$, and write $v:=\ell/L$.
We assume $v=O(1)$.
Let
\eq
\hN_\ell:=\sum_{x=1}^\ell c^*_x\,c_x,\quad
\hH_\ell:=\frac{1}{2}\sum_{x=1}^\ell\bigl\{e^{i\theta}\,c^*_x\,c_{x+1}
+e^{-i\theta}\,c^*_{x+1}\,c_{x}\bigr\}
\lb{HN}
\en
be the partial number operator and the partial energy operator, respectively, in the interval $\{1,\ldots,\ell\}$.
By explicit (and slightly nontrivial) calculations, whose details can be found in Appendix~\ref{s:fermiCal},  we see that
\eq
\bkt{\PG,\hN_\ell\,\PG}=v\,N,\quad
\bkt{\PG,(\hN_\ell)^2\,\PG}=(v\,N)^2+(v-v^2)\,N+O(\ell)
\lb{NN2}
\en
for any $\Gamma$.
Thus one has $\langle\hN_\ell\rangle_{\rm mc}=v\,N$ for any $E$ and $\DE$, and 
\eq
\bkt{\PG,(\hN_\ell- \langle\hN_\ell\rangle_{\rm mc})^2\,\PG}
=(v-v^2)\,N+O(\ell)=O(L).
\en
This means that $\hN_\ell$ is thermodynamically normal with respect to $\hH$ with the choice $\bar{N}_\ell=N=O(L)$ and $\zeta=O(1/L)\ll1$.

Similarly we can show that
\eqg
\bkt{\PG,\hH_\ell\,\PG}=v\,E_\Gamma\\
\bkt{\PG,(\hH_\ell)^2\,\PG}=(v\,E_\Gamma)^2+(v-v^2)\sum_{n\in\Gamma}(\epsilon_n)^2+O(\ell)
\lb{EE2}
\eng
for any $\Gamma$.
Then in the microcanonical ensemble in the range $[E,E+\DE]$, one has
\eq
\langle\hH_\ell\rangle_{\rm mc}=v\,E+O(\DE),
\en
and
\eqa
\bkt{\PG,(\hH_\ell- \langle\hH_\ell\rangle_{\rm mc})^2\,\PG}
&=(v-v^2)\sum_{n\in\Gamma}(\epsilon_n)^2+v\,(E_\Gamma-E)^2+O(\ell)
\ret&\le\,(v-v^2)\,E_\Gamma+O(\ell)=O(L),
\lb{Hln}
\ena
where we noted that\footnote{
This seemingly innocent manipulation relies essentially on the fact that the kinetic energy in the tight-binding model is bounded from above.
We expect that in the continuum model, where $\epsilon_n$ is not bounded, the quantity corresponding to $\hH_\ell$ is not normal.
This, we expect, is an artifact of an ideal gas.
} $|\epsilon_n|\le 1$.
Thus $\hH_\ell$ is thermodynamically normal with $\bar{H}_\ell=E=O(L)$ and $\zeta=O(1/L)\ll1$.

\subsubsection{System with general odd $L$}
\label{s:oddL}
Let us consider a system whose size $L$ is a (large) odd integer, but not necessarily a prime number.
In this case the energy eigenvalues $E_\Gamma$ may be degenerate.
Then in the long time average \rlb{Tinf}, off-diagonal terms $\bkt{\PG,(\hA-\Amc)^2\,\Phi_{\Gamma'}}$ with $\Gamma\ne\Gamma'$ but $E_\Gamma=E_{\Gamma'}$ do not vanish.

However Lemma 2 in Appendix~\ref{a:nond} implies that, for $\theta\in(0,\pi/N)$ except for a finite number of exceptional points, the degeneracy $E_\Gamma=E_{\Gamma'}$ for $\Gamma\ne\Gamma'$ is only possible when $\Gamma$ and $\Gamma'$ differ in more than two elements (i.e., $|\Gamma\backslash\Gamma'|=|\Gamma'\backslash\Gamma|>2$).
We shall show in Appendix~\ref{s:fermiCal} that for such $\Gamma$ and $\Gamma'$ one has
\eq
\bkt{\PG,\hN_\ell\,\Phi_{\Gamma'}}=0,\quad
\bkt{\PG,(\hN_\ell)^2\,\Phi_{\Gamma'}}=0
\lb{OD}
\en
and
\eq
\bkt{\PG,\hH_\ell\,\Phi_{\Gamma'}}=0,\quad
\bkt{\PG,(\hH_\ell)^2\,\Phi_{\Gamma'}}=0.
\en
Therefore we see that the unwanted off-diagonal elements vanish, and the equality in \rlb{Tinf} is valid as it is.
We have thus shown that both $\hN_\ell$ and $\hH_\ell$ exhibit the approach to equilibrium in the sense of our main theorem.

\subsection{Two identical system in a thermal contact}
\label{s:Two}
The third example is a simple model of two identical systems in thermal contact.
We show that the operator $\hH^{(1)}$, the Hamiltonian of one of the two systems, is thermodynamically normal when the ``volume'' of the system is large.
Then the main theorem guarantees that $\hH^{(1)}$ approaches its equilibrium value, even if one starts from a highly nonequilibrium initial state.
Thus the well known fact that ``two bodies in contact with each other will transfer heat until their temperatures are the same'' has been derived only from quantum time evolution (although in a highly artificial model).

We note that the present model is constructed so that our analysis can be carried out easily, and is hence extremely artificial.
We nevertheless hope that it somehow mimics basic structures of energy eigenvalues in more realistic systems.

Let us first describe a single system.
The energy eigenvalue is $n\,\ez$ with $n=1,2,\ldots$, where $\ez>0$ is a constant.
The energy level with $n\,\ez$ is $\Omega_n$ fold degenerate.
To mimic a healthy macroscopic system, the degeneracy is assumed to behave as
\eq
\Omega_n\simeq\exp[V\,s_n],
\en
where $V$ is the ``volume'' of the system, which will be made large.
The ``entropy'' $s_n>0$ is increasing in $n$, and satisfies the strict concavity
\eq
2\,s_n>s_{n-1}+s_{n+1}
\lb{sn}
\en
for any $n$.
An energy eigenstate with energy $n\,\ez$ is written as $\psi^{(1)}_{(n,j)}$ where $n=1,2,\ldots$, and $j=1,\ldots,\Omega_n$.
We denote the corresponding Hamiltonian as $\hH^{(1)}$.
Thus $\hH^{(1)}\,\psi^{(1)}_{(n,j)}=n\,\ez\,\psi^{(1)}_{(n,j)}$.

We prepare an exact copy of this system and denote its Hamiltonian and eigenstates as $\hH^{(2)}$ and $\psi^{(2)}_{(n',j')}$, respectively.
Thus $\hH^{(2)}\,\psi^{(2)}_{(n',j')}=n'\,\ez\,\psi^{(2)}_{(n',j')}$.

The total Hamiltonian of the system is given by\footnote{
Rigorously speaking, this equation should be $\hH=\hH^{(1)}\otimes\boldsymbol{1}+\boldsymbol{1}\otimes\hH^{(2)}+\varepsilon\,\hV$.
} $\hH=\hH^{(1)}+\hH^{(2)}+\varepsilon\,\hV$, where $\hV$ is the interaction that we shall define.

Let $m\ge2$ be an integer, and let $n,n'\in\{1,2,\ldots\}$ be such that $n+n'=m$.
Then any state $\psi^{(1)}_{(n,j)}\otimes\psi^{(2)}_{(n',j')}$ is an eigenstate of the noninteracting Hamiltonian $\hH^{(1)}+\hH^{(2)}$ with the eigenvalue $m\,\ez$.
The degeneracy of this eigenvalue is given by
\eq
\tilde{\Omega}_m=\sumtwo{n,n'\ge1}{(n+n'=m)}\Omega_n\,\Omega_{n'}
\lb{tOm}
\en
We denote by $\calH_m$ this $\tilde{\Omega}_m$-dimensional eigenspace.
The interaction $\hV$ is designed so that to leave each $\calH_m$ invariant, and mixes up the basis states within it.

Although it is likely that almost any ``generic'' interaction $\hV$ will do, we shall give a simple example of $\hV$ to be concrete (and constructive).
Fix $m$, and renumber the basis states $\psi^{(1)}_{(n,j)}\otimes\psi^{(2)}_{(n',j')}$ (with the restriction $n+n'=m$) of $\calH_m$ in an arbitrary manner\footnote{
For example, one can use the lexicographic ordering in $n$, $j$, and $j'$. 
} and call them $\tilde{\psi}_k$ with $k=1,\ldots,\tilde{\Omega}_m$.
Then we define
\eq
\bkt{\tilde{\psi}_k,\hV\,\tilde{\psi}_{k'}}=
\begin{cases}
1/2&\text{if $|k-k'|=1$}\\
0&\text{otherwise}.
\end{cases}
\lb{V}
\en
We do the same thing for all $m$ to define $\hV$ completely.

The interaction \rlb{V} lifts the degeneracy in $\calH_m$ completely,
and the normalized eigenstate is given by
\eq
\psi_{m,\ell}:=
\sqrt{\frac{2}{\tilde{\Omega}_m+1}}\,\sum_{k=1}^{\tilde{\Omega}_m}\sin\Bigl(\frac{\pi\,\ell}{\tilde{\Omega}_m+1}k\Bigr)\,\,\tilde{\psi}_k
\lb{phiml}
\en
where $\ell=1,\ldots,\tilde{\Omega}_m$.
The corresponding eigenvalue of the total Hamiltonian $\hH$ is
\eq
E_{m,\ell}=m\,\ez+\varepsilon\,\cos\Bigl(\frac{\pi\,\ell}{\tilde{\Omega}_m+1}\Bigr).
\en
Thus, by taking $\varepsilon\ll\ez$, we see that the eigenvalues of $\hH$ are nondegenerate.
This peculiar ``band structure'' of the eigenvalues of $\hH$ makes the analysis easy.
Although the band structure is just an artifact of the construction, we hope that the basic structure of the eigenstates \rlb{phiml}  mimics that in more realistic systems\footnote{
One can avoid too explicit band structures by considering (yet artificial) model in the line of \cite{unpub1}.
}.

Fix an even $m$, and consider the microcanonical ensemble for the energy interval $[m\,\ez-\varepsilon,m\,\ez+\varepsilon]$.
The ensemble consists precisely of the sates in $\calH_m$.

Let us focus on the observable $\hH^{(1)}$, the energy of one of the two systems.
Clearly its microcanonical average is $\langle\hH\rangle_{\rm mc}=m\,\ez$.
We shall show that $\hH^{(1)}$ is thermodynamically normal when $V$ is large.

Let $n$, $n'$ be such that $n+n'=m$.
The number of the basis states of $\calH_m$ of the form  $\psi^{(1)}_{(n,j)}\otimes\psi^{(2)}_{(n',j')}$ is $\Omega_n\,\Omega_{n'}\simeq \exp[V(s_n+s_{n'})]$.
Now write $n=(m/2)+r$ and $n'=(m/2)-r$.
From the concavity \rlb{sn}, it follows that the quantity $s_n+s_{n'}=s_{(m/2)+r}+s_{(m/2)-r}$ attains its maximum at $r=0$ and decreases monotonically as $r$ increases or decreases from 0\footnote{
{\em Proof:}\/ \rlb{sn} implies $s_n-s_{n-1}>s_{n+1}-s_n$.
By repeatedly using this, one finds $s_p-s_{p-1}>s_{q+1}-s_q$ for any $p\le q$.
This means $s_p+s_q>s_{p-1}+s_{q+1}$, which justifies the claim.
}.
We thus find
\eq
\frac{\Omega_{(m/2)+r}\,\Omega_{(m/2)-r}}{\Omega_{m/2}\,\Omega_{m/2}}
\le e^{-V\,\kappa_m},
\lb{OOOO}
\en
for any $|r|\ge1$, where $\kappa_m:=2s_{m/2}-s_{(m/2)-1}-s_{(m/2)+1}>0$.
This means that when $V$ is large the space $\calH_m$ mostly consists of the states $\psi^{(1)}_{(m/2,j)}\otimes\psi^{(2)}_{(m/2,j')}$.
Since the eigenstate \rlb{phiml} is an almost democratic linear combination of all the basis states, one readily finds that 
\eq
\bkt{\psi_{m,\ell},(\hH^{(1)})^2\,\psi_{m,\ell}}=(m\,\ez)^2\,\{1+O( e^{-V\,\kappa_m})\},
\en
which means
\eq
\bkt{\psi_{m,\ell},(\hH^{(1)}-\langle\hH^{(1)}\rangle_\mathrm{mc})^2\,\psi_{m,\ell}}=(m\,\ez)^2\,O( e^{-V\,\kappa_m}).
\en
Therefore the partial energy $\hH^{(1)}$ is thermodynamically normal with the choice $\bar{H}^{(1)}=m\,\ez$ and $\zeta=O( e^{-V\,\kappa_m})\ll1$.

We conclude that $\hH^{(1)}$ approaches its equilibrium value $m\,\ez$ when we start from an arbitrary initial state from $\calH_m$.
Note that $\calH_m$ contains such states where one system has much higher energy than the other.

Finally let us remark how the analysis of  Goldstein, Lebowitz, Mastrodonato, Tumulka, and Zangh\`\i\  \cite{GLMTZ09b} can be carried out in the present example.
If we identify $\calH_m$ above with $\HE$, the equilibrium Hilbert space $\calH_\mathrm{eq}$ of \cite{GLMTZ09b} is the one spanned by  the states $\psi^{(1)}_{(m/2,j)}\otimes\psi^{(2)}_{(m/2,j')}$ with $j,j'=1,\ldots,\Omega_{m/2}$.
Since $\mathrm{dim}\,\calH_m=\tilde{\Omega}_m$, and $\mathrm{dim}\,\calH_\mathrm{eq}=(\Omega_{m/2})^2$, the relations  \rlb{tOm} and \rlb{OOOO} justify \rlb{dH1} about the ratio of the dimensions of the spaces.
The desired thermodynamic normality \rlb{mynormal} of $\calH_\mathrm{eq}$ is then apparent from the expression \rlb{phiml} of the energy eigenstate.

\section{Discussions}
\label{s:discussion}

In the present note we have shown that the thermodynamic normality of a macroscopic observable guarantees that the measured value of the observable approaches the corresponding equilibrium value when one starts from  {\em any}\/ initial state which have energy concentrated around a certain macroscopic value.
This is a straightforward reinterpretation of the result of Goldstein, Lebowitz, Mastrodonato, Tumulka, and Zangh\`\i\ \cite{GLMTZ09b}.
We also presented three models in which the thermodynamic normality can be established for certain observables, thus providing concrete examples in which the ``approach to equilibrium'' is rigorously justified.

\bigskip

Here we followed  von Neumann \cite{vN29} and Goldstein, Lebowitz, Mastrodonato, Tumulka, and Zangh\`\i\  \cite{GLMTZ09b} in concentrating (except in section~\ref{s:most}) on results which hold for {\em any}\/ initial state from $\HE$.
That we allow basically an arbitrary initial state is a great advantage.
Once we have a concrete model satisfying the necessary criteria, then we can justify the use of equilibrium statistical mechanics without resorting to any probabilistic (or measure theoretic) arguments\footnote{
We still have to restrict ourselves to ``overwhelming majority of time'' in the  long interval $[0,T]$.
But this has no conceptually difficulty like ``probability'' that the initial pure state falls into a specific subspace of the Hilbert space.
}.
This makes the whole scenario extremely clear.

But, to be honest, we are still not sure how essential this point is.
It also sounds quite reasonable that results like those in section~\ref{s:most} or in \cite{Tasaki98,Reimann}, which hold for ``most'' initial states  from $\HE$, are sufficient from a ``practical'' point of view since the ``chance'' of preparing an inhibited initial state is negligible.

\bigskip

It must be seriously questioned whether the thermodynamic normality holds for realistic macroscopic observables in a realistic macroscopic system which exhibits the approach to equilibrium\footnote{
As we have remarked at the end of section~\ref{s:spin}, a class of systems with quenched randomness provide examples which lack the ability to relax to equilibrium by themselves.
Then the thermodynamic normality cannot be satisfied.
Here we are concerned (or worried) with those examples which do have the ability to relax but do not satisfy the thermodynamic normality.
}.
After all the thermodynamic normality is a very strong condition, and one might criticize that we are just assuming what we want.

For the moment all that we can hope to do is to study further examples and learn more about the nature of energy eigenstates in interacting macroscopic systems.
This is, of course, a formidably difficult task from the present standard of theoretical or mathematical physics.
Numerical results for the Bose-Hubbard model in \cite{BKL} may be interpreted as an indication that the thermodynamic normality fails in a non-random interacting quantum many-body system.

It may finally turn out that in general the strict thermodynamic normality does not hold in the sense that there exit certain exceptional energy eigenstates which considerably differ from the equilibrium state.
Then we will have to be satisfied with results, like those in section~\ref{s:most} or in \cite{Tasaki98,Reimann}, which hold for ``most'' initial states.

\bigskip
It is also interesting to look for sufficient conditions for the thermodynamic normality, and examine their validity.
For example, suppose that the Hamiltonian and hence the energy eigenstates are translation invariant, and the macroscopic observable $\hA$ is written as $\hA=\sum_{j=1}^N\hat{a}_j$, where each $\hat{a}_j$ is a translational copy of a local observable $\hat{a}_1$.
Then the assumptions that, for any $\alpha$ such that $E_\alpha\in[E,E+\DE]$,  the local expectation value satisfies
\eq
\bkt{\psi_\alpha,\hat{a}_1\,\psi_\alpha}\simeq a,
\en
where $a$ is a constant, and the truncated correlation
\eq
\bkt{\psi_\alpha,\hat{a}_j\,\hat{a}_{j'}\,\psi_\alpha}-\bkt{\psi_\alpha,\hat{a}_j\,\psi_\alpha}\,\bkt{\psi_\alpha,\hat{a}_{j'}\,\psi_\alpha}
\en
decays fast enough in the distance between the supports of $\hat{a}_j$ and $\hat{a}_{j'}$ are enough to guarantee the thermodynamic normality of $\hA$.

\bigskip
When the system exhibits a phase coexistence, the thermodynamic normality cannot be satisfied as it is.
To treat such a situation, it seems that we need to consider a projection operator explicitly.
Suppose, for simplicity, that two phases + and $-$ coexist at a macroscopic energy $E$.
Then the microcanonical average can be decomposed as
\eq
\sbkt{\hA}_\mathrm{mc}\simeq\eta\,\sbkt{\hA}_\mathrm{mc}^++(1-\eta)\,\sbkt{\hA}_\mathrm{mc}^-,
\en
where $\sbkt{\cdots}_\mathrm{mc}^\pm$ are expectations in each phase, and $\eta\in(0,1)$ is a constant.
We then define $\hat{P}_{\hat{A}}$ as the projection operator onto the subspace where $\hA$ takes a value close to $\sbkt{\hA}_\mathrm{mc}^+$ or $\sbkt{\hA}_\mathrm{mc}^-$.
$\hat{P}_{\hat{A}}$ is thus a sum of two projections\footnote{
It may be nontrivial to define a suitable projection operator in a general situation where, for example, spatial phase coexistence (separated by a domain wall) takes place.
}.
Then our notion of thermodynamic normality should be replaced by
\eq
\bkt{\psi_\alpha,\hat{P}_{\hat{A}}\,\psi_\alpha}\simeq1
\quad\text{for any $\alpha$ such that  $E_\alpha\in[E,E+\DE]$}.
\en
But such a condition involving a projection operator brings us back to the formulation of Goldstein, Lebowitz, Mastrodonato, Tumulka, and Zangh\`\i\  \cite{GLMTZ09b}. See section~\ref{s:rel}.

\appendix
\section{Some details of the free fermion model}
We supply some technical details necessary to get the results of section~\ref{s:fermi}.

\subsection{Proof of nondegeneracy}
\label{a:nond}
Throughout the present subsection we make the $\theta$ dependence of the energies explicit and write $\ent=\cos(2\pi(n/L)+\theta)$, and $E_\Gamma(\theta)=\sum_{n\in\Gamma}\ent$.
We also define and use the complex constant $\zeta:=\exp[(2\pi/L)i]$.

\bigskip

The first lemma states that the energy eigenvalues have no degeneracy for most $\theta$ provided that the system size $L$ is a prime number.

\bn
{\bf Lemma 1:} 
Let $L>2$ be a prime number, and fix any particle number $N\le L$.
Then the following is true for $\theta\in(0,\pi/L)$ except for a finite number of points.
For any $\Gamma,\Gamma'\subset\oL$ such that $|\Gamma|=|\Gamma'|=N$ and $\Gamma\ne\Gamma'$, one has $E_\Gamma(\theta)\ne E_{\Gamma'}(\theta)$.

\bn
{\em Proof:}\/
For each $\Gamma\subset\oL$, let $z(\Gamma):=\sum_{n\in\Gamma}\zeta^n$.
We shall show below that $z(\Gamma)\ne0$ for any $\Gamma$ such that $|\Gamma|<L$, and $z(\Gamma)\ne z(\Gamma')$ for $\Gamma$ and $\Gamma'$ such that $|\Gamma|=|\Gamma'|<L/2$ and $\Gamma\ne\Gamma'$.

Let us prove the Lemma when $N<L/2$.
The case $N>L/2$ then follows by noting that $\sum_{n=1}^L\ent=0$, and hence $E_\Gamma=-E_{\oL\backslash\Gamma}$.

Consider all the subsets $\Gamma\subset\oL$ with $|\Gamma|=N$.
From what we stated above, $z(\Gamma)$ corresponding to these subsets are all distinct and nonzero.
Denote by $P_N$ the set of these $\binom{L}{N}$ distinct and nonzero points in the complex plane.
By noting that $E_\Gamma(\theta)=\Re[z(\Gamma)\,e^{i\theta}]$, we see that the energy eigenvalues of the $N$ particle system is obtained by first rotating $P_N$ around the origin by $\theta$, and then projecting the resulting sets to the real axis.
Since there are only finite number of points, the projected images may overlap but only for finite ``unlucky'' values of $\theta$.
This proves the desired nondegeneracy.

It remains to show the claim about the properties of $z(\Gamma)$.
Here we shall make use of the classical result by Gauss known as ``the irreducibility of the cyclotomic polynomials of prime index" (see, for example, Chapter 12, Section 3 of \cite {Tignol} or Chapter 13, Section 2 of \cite{IR}).
It implies that the $N-1$ complex numbers $\zeta$, $\zeta^2$, $\ldots$, $\zeta^{N-1}$ are rationally independent.
More precisely, if $\sum_{n=1}^{N-1}m_n\,\zeta^n=0$ with integers $m_1,\ldots,m_{N-1}$, one inevitably has $m_1=m_2=\cdots=m_{N-1}=0$.

To show the first claim, we choose  and fix $n_0\in\oL\backslash\Gamma$.
Such $n_0$ exists whenever $|\Gamma|<L$.
Then $\zeta^{-n_0}\,z(\Gamma)=\sum_{n\in\Gamma}\zeta^{n-n_0}$ is rewritten as $\sum_{n=1}^{N-1}m_n\,\zeta^n$ with $(m_1,m_2,\ldots,m_{N-1})\ne(0,0,\ldots,0)$.
From the rational independence, we see $z(\Gamma)\ne0$.

To show the second claim, we choose and fix $n_0\in\oL\backslash(\Gamma\cup\Gamma')$.
Such a $n_0$ exists since $|\Gamma|=|\Gamma'|<L/2$.
Again $\zeta^{-n_0}\,\{z(\Gamma)-z(\Gamma')\}=\sum_{n\in\Gamma}\zeta^{n-n_0}-\sum_{n\in\Gamma'}\zeta^{n-n_0}$ is rewritten as $\sum_{n=1}^{N-1}m_n\,\zeta^n$ with $(m_1,m_2,\ldots,m_{N-1})\ne(0,0,\ldots,0)$.
From the rational independence, we see $z(\Gamma)-z(\Gamma')\ne0$.~\qedm

\bigskip

The second lemma states the property of the energy eigenvalues used in section~\ref{s:oddL}.

\bn
{\bf Lemma 2:} 
When $L$ is an odd positive integer, the following is true for $\theta\in(0,\pi/L)$ except for finite number of points.
For any $n,n',m,m'\in\{1,2,\ldots,N\}$ with $n<n'$ and $m<m'$, and $(n,n')\ne(m,m')$, one has
\eq
\epsilon_{n}(\theta)+\epsilon_{n'}(\theta)\ne\epsilon_{m}(\theta)+\epsilon_{m'}(\theta).
\en

\bn
{\em Proof:}\/
Note that the statement of Lemma 2 is that of Lemma 1 restricted to $N=2$.
Thus it suffices to show that the set $P_2$ consists of $L(L-1)/2$ distinct nonzero points.
This is easily proved by the following geometric consideration (we suggest the reader to draw simple figures while reading the following).
Since $L$ is odd, it is trivial that $\zeta^n+\zeta^{n'}\ne0$.
To show the distinctness, fix $n$, $n'$, and examine the relation $\zeta^n+\zeta^{n'}=u+v$ where $u,v,\in\mathbb{C}$ such that $|u|=|v|=1$.
Since the simultaneous equations $|u|=1$, $|\zeta^n+\zeta^{n'}-u|=1$ for $u$ has at most two solutions (corresponding to the two intersection points of two circles), we find that only $u, v$ satisfying the above conditions are $(u,v)=(\zeta^n,\zeta^{n'})$ or $(u,v)=(\zeta^{n'},\zeta^{n})$.~\qedm

\subsection{Evaluation of the expectation values}
\label{s:fermiCal}
We describe calculations needed to get the results in sections \ref{s:fermiN} and \ref{s:oddL}.
Although the required techniques are standard, some results require nontrivial estimates.
To help the readers who do not have experiences in many-body quantum systems, we also explain standard estimates in some detail.
For a more systematic exposition, see, for example, \cite{Hub}.

For $\ell<L$ and $n\in\oL$, define the ``incomplete'' creation operator
\eq
b^*_n:=\frac{1}{\sqrt{L}}\sum_{x=1}^\ell \exp\Bigl[i\,\dfrac{2\pi\,n}{L}\,x\Bigr]\,c^*_x,
\en
which corresponds to $a^*_n$ defined in \rlb{as}.
From the basic anticommutation relations \rlb{cac1}, we find
\eq
\{a_n,b^*_n\}=\frac{\ell}{L}=v,
\lb{ab1}
\en
for any $n$.
It is important to note that, for $n\ne n'$, the anticommutator
\eq
\{a_{n},b^*_{n'}\}=
\frac{1}{L}\sum_{x=1}^\ell\exp\Bigl[i\,\dfrac{2\pi\,(n'-n)}{L}\,x\Bigr]
=:w_{n,n'}
\lb{ab2}
\en
is nonvanishing  because the sum in $x$ is incomplete.

\bigskip
Let us describe how to calculate the quantities involving $\hN_\ell$.

From \rlb{cac1}, we get the commutation relations
\eq
[c^*_x\,c_y,c^*_z]=\delta_{y,z}\,c^*_x
\lb{ccc}
\en
for any $x,y,z\in\oL$.
This relation and the definition \rlb{HN} implies that 
\eq
[\hN_\ell,a^*_n]=b^*_n, \quad [\hN_\ell,b^*_n]=b^*_n,
\lb{Na}
\en
which will be useful.

Take $\Gamma\subset\oL$ with $|\Gamma|=N$, and write it as $\Gamma=\{n_1,n_2,\ldots,n_N\}$ so that $n_{j}<n_{j+1}$.
By using the definition \rlb{NG} of $\PG$, the relation $\hN_\ell\,\Pv=0$, the commutation relation \rlb{Na}, and the relation $\{a^*_n,b^*_{n'}\}=0$, we find
\eqa
\hN_\ell\,\PG&=
\hN_\ell\,a^*_{n_1}\cdots a^*_{n_N}\,\Pv
\ret&
=
[\hN_\ell,a^*_{n_1}\cdots a^*_{n_N}]\,\Pv
\ret
&=
\sum_{j=1}^Na^*_{n_1}\cdots a^*_{n_{j-1}}\,[\hN_\ell,a^*_{n_j}]\,a^*_{n_{j+1}}\cdots a^*_{n_{N}}\,\Pv
\ret
&=
\sum_{j=1}^Na^*_{n_1}\cdots a^*_{n_{j-1}}\,b^*_{n_j}\,a^*_{n_{j+1}}\cdots a^*_{n_{N}}\,\Pv
\ret
&=
\sum_{j=1}^N(-1)^{j-1}\,b^*_{n_j}\,a^*_{n_1}\cdots a^*_{n_{j-1}}\,a^*_{n_{j+1}}\cdots a^*_{n_{N}}\,\Pv
\ret
&=\sum_{n\in\Gamma}\sigma(\Gamma;n)\,b^*_n\,\Bigl(\prod_{n'\in\Gamma\backslash\{n\}}a^*_{n'}\Bigr)\Pv,
\lb{NPG}
\ena
where $\sigma(\Gamma;n)=(-1)^{j-1}$ is the fermion sign.
To calculate the expectation value, we rewrite $\PG$ similarly as 
\eq
\PG=\sigma(\Gamma;n)\,a^*_n\,\Bigl(\prod_{n'\in\Gamma\backslash\{n\}}a^*_{n'}\Bigr)\Pv,
\en
for each $n\in\Gamma$.
Then we have
\eqa
\bkt{\PG,\hN_\ell\,\PG}&=
\sum_{n\in\Gamma}\bkt{
a^*_n\,\Bigl(\prod_{n'\in\Gamma\backslash\{n\}}a^*_{n'}\Bigr)\Pv,
b^*_n\,\Bigl(\prod_{n'\in\Gamma\backslash\{n\}}a^*_{n'}\Bigr)\Pv
}
\ret
&=\sum_{n\in\Gamma}\bkt{
\Bigl(\prod_{n'\in\Gamma\backslash\{n\}}a^*_{n'}\Bigr)\Pv,
a_n\,b^*_n\,\Bigl(\prod_{n'\in\Gamma\backslash\{n\}}a^*_{n'}\Bigr)\Pv
}
\ret
&=\sum_{n\in\Gamma}v\,\bkt{
\Bigl(\prod_{n'\in\Gamma\backslash\{n\}}a^*_{n'}\Bigr)\Pv,
\Bigl(\prod_{n'\in\Gamma\backslash\{n\}}a^*_{n'}\Bigr)\Pv
}
\ret
&=v\,N,
\ena
where we used \rlb{ab1}.
We have also noted that 
$a_n\,\Bigl(\prod_{n'\in\Gamma\backslash\{n\}}a^*_{n'}\Bigr)\Pv$
\newline$=(-1)^N\,\Bigl(\prod_{n'\in\Gamma\backslash\{n\}}a^*_{n'}\Bigr)\,a_n\,\Pv=0$.

To evaluate the expectation value of $(\hN_\ell)^2$, we proceed as in \rlb{NPG} to get
\eq
(\hN_\ell)^2\,\PG
=\sum_{n\in\Gamma}\sigma(\Gamma;n)\,b^*_n\,\Bigl(\prod_{n'\in\Gamma\backslash\{n\}}a^*_{n'}\Bigr)\Pv,
+2\sumtwo{n,n'\in\Gamma}{(n<n')}\sigma(\Gamma;n,n')
\,b^*_n\,b^*_{n'}\,\Bigl(\prod_{n''\in\Gamma\backslash\{n,n'\}}a^*_{n''}\Bigr)\Pv,
\lb{NNPG}
\en
where $\sigma(\Gamma;n,n')=\pm1$ is an appropriate fermion sign.
Then we have
\eqa
\bkt{\PG,(\hN_\ell)^2\,\PG}=&
\sum_{n\in\Gamma}\bkt{
a^*_n\,\Bigl(\prod_{n'\in\Gamma\backslash\{n\}}a^*_{n'}\Bigr)\Pv,
b^*_n\,\Bigl(\prod_{n'\in\Gamma\backslash\{n\}}a^*_{n'}\Bigr)\Pv
}
\ret&
+2\sumtwo{n,n'\in\Gamma}{(n<n')}\bkt{
a^*_n\,a^*_{n'}\,\Bigl(\prod_{n''\in\Gamma\backslash\{n,n'\}}a^*_{n''}\Bigr)\Pv,
b^*_n\,b^*_{n'}\,\Bigl(\prod_{n''\in\Gamma\backslash\{n,n'\}}a^*_{n''}\Bigr)\Pv
}
\ret
&=v\,N
+2\sumtwo{n,n'\in\Gamma}{(n<n')}\bkt{
\Bigl(\prod_{n''\in\Gamma\backslash\{n,n'\}}a^*_{n''}\Bigr)\Pv,
a_{n'}\,a_n\,b^*_n\,b^*_{n'}\,\Bigl(\prod_{n''\in\Gamma\backslash\{n,n'\}}a^*_{n''}\Bigr)\Pv
}.
\lb{ENN}
\ena
By using the anticommutation relations \rlb{ab1}, \rlb{ab2} repeatedly we get
\eq
a_{n'}\,a_n\,b^*_n\,b^*_{n'}
=v^2-|w_{n,n'}|^2-v\,b^*_{n'}\,a_{n'}+w_{n,n'}\,b^*_n\,a_{n'}+a_{n'}\,b^*_n\,b^*_{n'}\,a_n.
\lb{aabb}
\en
Since the non-constant parts acting on $\Bigl(\prod_{n''\in\Gamma\backslash\{n,n'\}}a^*_{n''}\Bigr)\Pv$ give 0, we conclude
\eq
\bkt{\PG,(\hN_\ell)^2\,\PG}=v\,N+2\sumtwo{n,n'\in\Gamma}{(n<n')}(v^2-|w_{n,n'}|^2)
=(v\,N)^2+(v-v^2)\,N-2\sumtwo{n,n'\in\Gamma}{(n<n')}|w_{n,n'}|^2.
\lb{ENN2}
\en
Since $(v\,N)^2+(v-v^2)\,N$ is what one gets from an independent distribution of particles, the final sum is a nontrivial correction that we shall now control.

Fix $n$ and $n'$ such that $n<n'$.
Note that the oscillating factor $\exp[i\,\{2\pi(n'-n)/L\}\,x]$ in \rlb{ab2} has the wave length $L/(n'-n)$.
When we sum this oscillating factor over an interval which is an integer multiple of the wave length, the oscillation perfectly cancels out and the result vanishes.
This leads us to a simple upper bound
\eq
\left|\sum_{x=1}^\ell\exp\Bigl[i\,\dfrac{2\pi\,(n'-n)}{L}\,x\Bigr]\right|\le\frac{L}{n'-n},
\en
which means 
\eq
|w_{n,n'}|^2\le\frac{1}{(n'-n)^2}.
\en
For $n$, $n'$ such that $L/(n'-n)\le\ell$, we use this upper bound.
For the remaining  $n$, $n'$ such that $L/(n'-n)>\ell$, we use the trivial bound $\bigl|\,\exp[i\,\{2\pi(n'-n)/L\}\,x]\,\bigr|\le1$ which gives
\eq
|w_{n,n'}|^2\le\left(\frac{\ell}{L}\right)^2=v^2.
\en
By replacing the range $n,n'\in\Gamma$ by $n,n'\in\oL$, we bound the correction term as
\eqa
\sumtwo{n,n'\in\Gamma}{(n<n')}|w_{n,n'}|^2&\le\sumtwo{n,n'\in\oL}{(n<n')}|w_{n,n'}|^2
\ret
&=\sumtwo{n,n'\in\oL}{(n'-n\ge L/\ell)}|w_{n,n'}|^2
+\sumtwo{n,n'\in\oL}{(L/\ell>n'-n>0)}|w_{n,n'}|^2
\ret&
\le
\sumtwo{n,n'\in\oL}{(n'-n\ge L/\ell)}\frac{1}{(n'-n)^2}
+\sumtwo{n,n'\in\oL}{(L/\ell>n'-n>0)}v^2
\ret&\le
L\int_{L/\ell}^\infty dx\,\frac{1}{x^2}+L\,\frac{L}{\ell}\,v^2=2\ell
\ena
Substituting this back to \rlb{ENN2}, we get the desired estimate
\eq
\bkt{\PG,(\hN_\ell)^2\,\PG}=(v\,N)^2+(v-v^2)\,N+O(\ell).
\en

Finally we briefly discuss the derivation of \rlb{OD} about the off-diagonal matrix elements.
Let $\Gamma$ and $\Gamma'$ be such that $|\Gamma|=|\Gamma'|=N$, $\Gamma\ne\Gamma'$, and $E_\Gamma=E_{\Gamma'}$.
Then as in \rlb{ENN}, we get
\eqa
\bkt{\Phi_{\Gamma'},(\hN_\ell)^2\,\PG}=&
\sum_{n\in\Gamma}\bkt{
\Bigl(\prod_{n'\in\Gamma'\backslash\{m\}}a^*_{n'}\Bigr)\Pv,
a_m\,b^*_n\,\Bigl(\prod_{n'\in\Gamma\backslash\{n\}}a^*_{n'}\Bigr)\Pv
}
\ret&
+2\sumtwo{n,n'\in\Gamma}{(n<n')}\bkt{
\Bigl(\prod_{n''\in\Gamma'\backslash\{m,m'\}}a^*_{n''}\Bigr)\Pv,
a_{m'}\,a_m\,b^*_n\,b^*_{n'}\,\Bigl(\prod_{n''\in\Gamma\backslash\{n,n'\}}a^*_{n''}\Bigr)\Pv
},
\lb{ENN3}
\ena
where, in the first sum, $m$ is chosen for each $n$ so that $m\not\in\Gamma\backslash\{n\}$.
Likewise, in the second sum, $m$, $m'$ are chosen for each $n$, $n'$ so that $m,m'\not\in\Gamma\backslash\{n,n'\}$.
Then we use the anticommutation relations \rlb{ab1}, \rlb{ab2} to re-order $a_m\,b^*_n$ and $a_{m'}\,a_m\,b^*_n\,b^*_{n'}$ as in \rlb{aabb} to get
\eqa
\bkt{\Phi_{\Gamma'},(\hN_\ell)^2\,\PG}=&
\sum_{n\in\Gamma}{\rm const.}\,\bkt{
\Bigl(\prod_{n'\in\Gamma'\backslash\{m\}}a^*_{n'}\Bigr)\Pv,
\Bigl(\prod_{n'\in\Gamma\backslash\{n\}}a^*_{n'}\Bigr)\Pv
}
\ret&
+\sumtwo{n,n'\in\Gamma}{(n<n')}{\rm const.}\,\bkt{
\Bigl(\prod_{n''\in\Gamma'\backslash\{m,m'\}}a^*_{n''}\Bigr)\Pv,
\Bigl(\prod_{n''\in\Gamma\backslash\{n,n'\}}a^*_{n''}\Bigr)\Pv
}.
\lb{ENN4}
\ena
But the right-hand side is vanishing since we know (from Lemma 2 in Appendix~\ref{a:nond}) that
\eq
\Gamma\backslash\{n\}\ne\Gamma'\backslash\{m\},\quad
\Gamma\backslash\{n,n'\}\ne\Gamma'\backslash\{m,m'\}
\en
for any choice of $n,n'\in\Gamma$ and $m,m'\in\Gamma'$.

\bigskip
The calculation of the quantities involving $\hH_\ell$ proceeds in essentially the same manner, except for some minor complications.
Let us make a few remarks about the differences.

From the basic commutation relations \rlb{ccc}, we find
\eq
[\hH_\ell,c^*_x]=
\begin{cases}
\frac{1}{2}\,e^{-i\theta}\,c^*_2&\text{if $x=1$}\\
\frac{1}{2}\,(e^{i\theta}\,c^*_{x-1}+e^{-i\theta}\,c^*_{x+1})&\text{if $x=2,\ldots,\ell$}\\
\frac{1}{2}\,e^{i\theta}\,c^*_{\ell}&\text{if $x=\ell+1$},
\end{cases}
\en
which implies
\eq
[\hH_\ell,a^*_n]=\epsilon_n\,b^*_n+e^{-i(\theta+k)}\,\frac{1}{2\sqrt{L}}
\bigl\{
e^{ik(\ell+1)}c^*_{\ell+1}-e^{ik}c^*_1
\bigr\}=:d^*_n,
\lb{Ha}
\en
where we wrote $k=2\pi n/L$.
Comparing with the corresponding relation \rlb{Na} for the partial number operator, we see that there appear extra boundary terms.
Further commutation gives
\eq
[\hH_\ell,d^*_n]=\epsilon_n\,b^*_n+\frac{1}{\sqrt{L}}\,(\text{finite number of terms})=:f^*_n.
\en
One then repeats the same calculations as above, carefully replacing $b^*_n$ by $d^*_n$ or $f^*_n$, and computing appropriate commutation relations.

\bigskip
It is a pleasure to thank Shelly Goldstein and Joel Lebowitz for useful discussions and comments, Shin-ichi Sasa for bringing my attention to \cite{GLTZ10,vN29} and for useful discussions, and David Huse and Giulio Biroli for useful comments on the manuscript.
I also thank Shin Nakano for valuable instruction concerning the result of Gauss used in the Appendix~\ref{a:nond}.

\end{document}